\documentclass[varenna]{cimento}

\providecommand{\sorthelp}[1]{}

\newcommand{\mytilde}{\raise.17ex\hbox{$\scriptstyle\mathtt{\sim}$}}

\def\spose#1{\hbox to 0pt{#1\hss}}
\def\simlt{\mathrel{\spose{\lower 3pt\hbox{$\mathchar"218$}}
     \raise 2.0pt\hbox{$\mathchar"13C$}}}
\def\simgt{\mathrel{\spose{\lower 3pt\hbox{$\mathchar"218$}}
     \raise 2.0pt\hbox{$\mathchar"13E$}}}

\usepackage{graphicx}  % got figures? uncomment this

\renewcommand{\thefootnote}{\arabic{footnote}}

\title{The Cosmic Neutrino Background}

\author{Douglas Scott\thanks{dscott@phas.ubc.ca}}

\institute{Dept.\ of Physics \& Astronomy, Univ.\ of British Columbia,
 Vancouver, Canada}

\begin{document}

\maketitle

\begin{abstract}
The cosmic neutrino background is like the cosmic microwave background, but
less photon-y and more neutrino-ey.  The CNB is also less talked about than the
CMB, mostly because it's nearly impossible to detect directly.  But if it
could be detected, it would be interesting in several ways that are
discussed.
\end{abstract}

\section{\boldmath$\nu$ overview (or `news')}
\label{sec:summary}
Let's start by giving an overview of the relevant properties
of neutrinos:

\begin{itemize}
  \item neutrinos are hard to detect;
  \item they come in 3 flavours, which can mix;
  \item we don't know why the flavour and mass states are so messed up (unlike
  the situation for quarks);
  \item we don't know if they're their own anti-particles or not (Majorana
  versus Dirac);
  \item we don't know their masses, or even if they have the ``normal
  hierarchy'' ($m_3\gg m_2\simgt m_1$) or an ``inverted hierarchy''
  ($m_2\simgt m_1\gg m_3$).
\end{itemize}

And let's also summarise some information for cosmological neutrinos
(making informal comments as notes among the references at the end,
like this one \cite{summary}):
%\footnote[42]{thing}

\begin{itemize}
  \item cosmological neutrinos are {\it really\/} hard to detect;
  \item there are about 340 for every cm$^3$ of the Universe
  (making them the second most common particle, after photons);
  \item if we could detect them, they'd tell us interesting things;
  \item the sum of neutrino masses is expected to be the 7th (or 8th)
  cosmological parameter.
\end{itemize}

This last point is why many cosmologists are currently excited about neutrinos.
There is a very strong expectation that soon it will be possible to make a
measurement of the sum of the neutrino masses, $\sum m_\nu$, because of the
effects of the cosmic neutrino background (CNB or C$\nu$B) on cosmological
power spectra, after the neutrinos become non-relativistic.
The standard cosmological model, also known as $\Lambda$CDM
(or ``cosmological-constant-dominated, cold dark matter''),
has six basic parameters, or seven if we include the amplitude of the photon
background.  The mass density of neutrinos in the CNB will be one additional
parameter.

For a good summary of neutrinos in cosmology, I recommend the review article
by Julien Lesgourgues and Licia Verde in the ``Particle Data Book'' (PDG)
\cite{LesgourguesVerde}.  What follows will be a broad discussion of the
cosmic neutrino background (CNB or C$\nu$B), from the perspective of someone
who has worked on the photon background, and would love to live in a universe
where it was routine to extract information from the CNB, just as it is now for
the CMB.  This will not be a particle-physics-oriented overview -- for
more on the particle properties of neutrinos I would suggest reading the
PDG review by Maria Gonzalez-Garcia and Masashi
Yokoyama \cite{Gonzalez-GarciaYokoyama}).  However, it will be important to
consider how neutrinos differ from photons, which of course involves
some particle physics \cite{simgt}.

\section{\boldmath$\nu$ cosmic background}
\label{sec:what}
As already stated, neutrinos are hard to detect, even if they're produced
copiously in the early Universe and other sources.
The Sun makes so many neutrinos that even at
the distance of the Earth, something like $10^{15}$ of them enter your body
every second~\cite{minimise}, and the same number also exit your body every
second.  There's a small chance that one of them might interact with you in
your lifetime.  Human bodies are of course not very sensitive detectors,
and we can do much better with a tailor-made experiment, but
nevertheless, the numbers detected are only measured in the tens per day.
Solar neutrinos are typically at energies in the MeV range.  What we're
focussing on in this article are the CNB neutrinos, which are also at
``em-ee-vee'' energies, but now it's meV rather than MeV.  Such low-energy
neutrinos basically don't interact with anything \cite{undetectable}. 

The neutrinos of the CNB were made at very early times, before the weak
interactions froze out.  We can calculate their number density today, which
turns out to be about $340\,{\rm cm}^{-3}$ for all flavours together
(and including anti-neutrinos).
That's only slightly less than the value $411\,{\rm cm}^{-3}$ for the photons
of the CMB.  This makes neutrinos the second most populous particle in the
Universe (if we don't count gravitons from inflation, at least \cite{Page2016}).
In terms of mass density,
the fraction of the cosmological critical density in massive neutrinos is
\begin{equation}
\Omega_\nu={\sum m_\nu\over 93.14\,h^2\,{\rm eV}},
\end{equation}
where the Hubble parameter is
$H_0=100\,h\,{\rm km}\,{\rm s}^{-1}\,{\rm Mpc}^{-1}$ and $h^2\simeq0.5$.
If we assume the minimal value for the sum of the neutrino masses (about
$0.06\,$eV, which
comes from adopting the normal hierarchy and using estimates of the
mass-squared differences from neutrino oscillation measurements), then we find
that $\Omega_\nu\simeq0.0014$.  This can be compared with the contributions
from baryons and cold dark matter, $\Omega_{\rm b}=0.049$ and
$\Omega_{\rm c}=0.261$, respectively \cite{planck2016-l06,Tristram2023}.

On the other hand, the estimated cosmic density of stars in an average
piece of the Universe is $\Omega_\ast\simeq0.0015$ \cite{FukPee2004}.  This
means that $\Omega_\nu\simeq\Omega_\ast$ \cite{conspiracies,mnemonics},
which is of course just a coincidence \cite{coincidence1},
but the reason to point it out is that it emphasises that the
neutrinos are {\it not\/} negligible, and {\it some of the dark matter consists
of neutrinos}.  It has been a pretty good approximation up until now
to ignore the effects of neutrino mass on structure formation, but as
measurements become more precise, then eventually we're not going to be able to
neglect the neutrinos.  We're now at the point where the
neutrinos need to be taken into account for the next generation of surveys.

Another way of saying the same thing is that if the neutrinos were a lot
more massive, then they would be the ideal dark matter candidates.  They are
already known to exist, are everywhere in the Cosmos and have no
electromagnetic interactions.  In some ways it's unfortunate that nature didn't
give us more massive neutrinos -- but we now know that their masses are
small enough that the fraction of dark matter that is neutrinos is almost
(but not completely) negligible and instead we need the dark matter to be
an {\it unknown\/} particle that also has very weak interactions \cite{CDM}.

\section{\boldmath$\nu$ background versus photon backgrounds}
\label{sec:photons}
\begin{figure}[htbp!]
\includegraphics[width=1.0\textwidth]{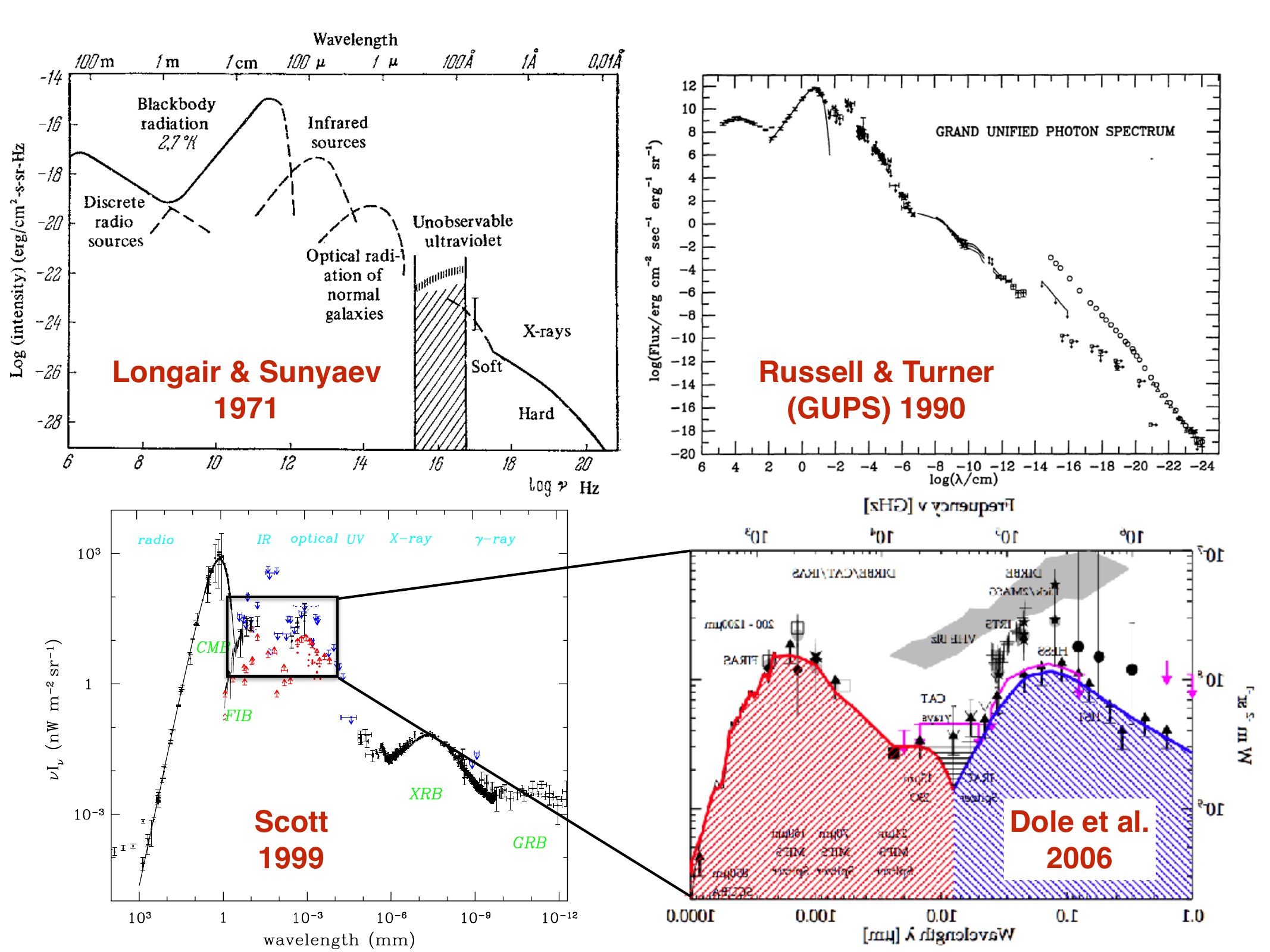}
\caption{Collection of historical compilations of observational constraints on
the extragalactic photon background.  The lower-left panel multiplies the
vertical axis by one more power of frequency, to give a better comparison of
the energy densities in various components.  The lower-right panel focuses on
the far-IR and optical backgrounds coming from stars.}
\label{fig:GUPS}
\end{figure}

Let's delve a bit more deeply into the CNB, by first having a closer look
at the cosmic photon backgrounds.  There's a long history of considering the
measurement and modelling of the emission from the \textit{extragalactic}
Universe, spanning all wavelengths in
the electromagnetic spectrum.  Figure~\ref{fig:GUPS} shows some examples of
reviews of this topic.  The earliest is probably that of Longair \& Sunyaev
in 1971 \cite{LongairSunyaev}, who, despite having very limited empirical
data at that time, got things astonishingly correct!  A significant advance
was made in 1990, when Ressell \& Turner presented the ``Grand Unified Photon
Spectrum'' \cite{RessellTurner}.  This compilation was updated (by me) and
published in conference proceedings in 1999 \cite{Scott99A,Scott99B,Scott99C},
as well as in Chapter~26 of ``Allen's Astrophysical Quantities'' \cite{AQ}.
This built on Ted Ressell's use of upper limits \cite{SM} to also include
lower limits from source counts, in order to bracket the measurements.  A more
detailed review of constraints on the infrared (IR) part of the background
was published by Dole, Lagache, Puget and collaborators in several articles
around 2006 \cite{Lagache2005,Dole2006} -- the last panel in
Fig.~\ref{fig:GUPS} shows the IR and optical parts of the background
\cite{flipped}.

\vspace{0.5cm}

The largest fraction of the total photon energy density is the CMB, thermal
radiation from the early Universe.
The optical and near-IR portion is dominated by direct starlight, while the
far-IR part traces the starlight absorbed and reradiated by interstellar dust.
Higher energy photons come mostly from active galactic nuclei (AGN) of various
types.
In general, the extragalactic photon background intensity that we measure is
an integral of the history of the photon luminosity density, ${\cal L}(z)$
of sources over all redshifts:
\begin{equation}
I= \left({c\over4\pi}\right) \int_0^\infty {\cal L}(z)
 \left|{dt\over dz} \right| {dz\over 1+z}.
\end{equation}
Thus, the background tells us about the history of star formation,
heavy-element production, black hole accretion, etc.

\begin{figure}[htbp!]
\includegraphics[width=1.0\textwidth]{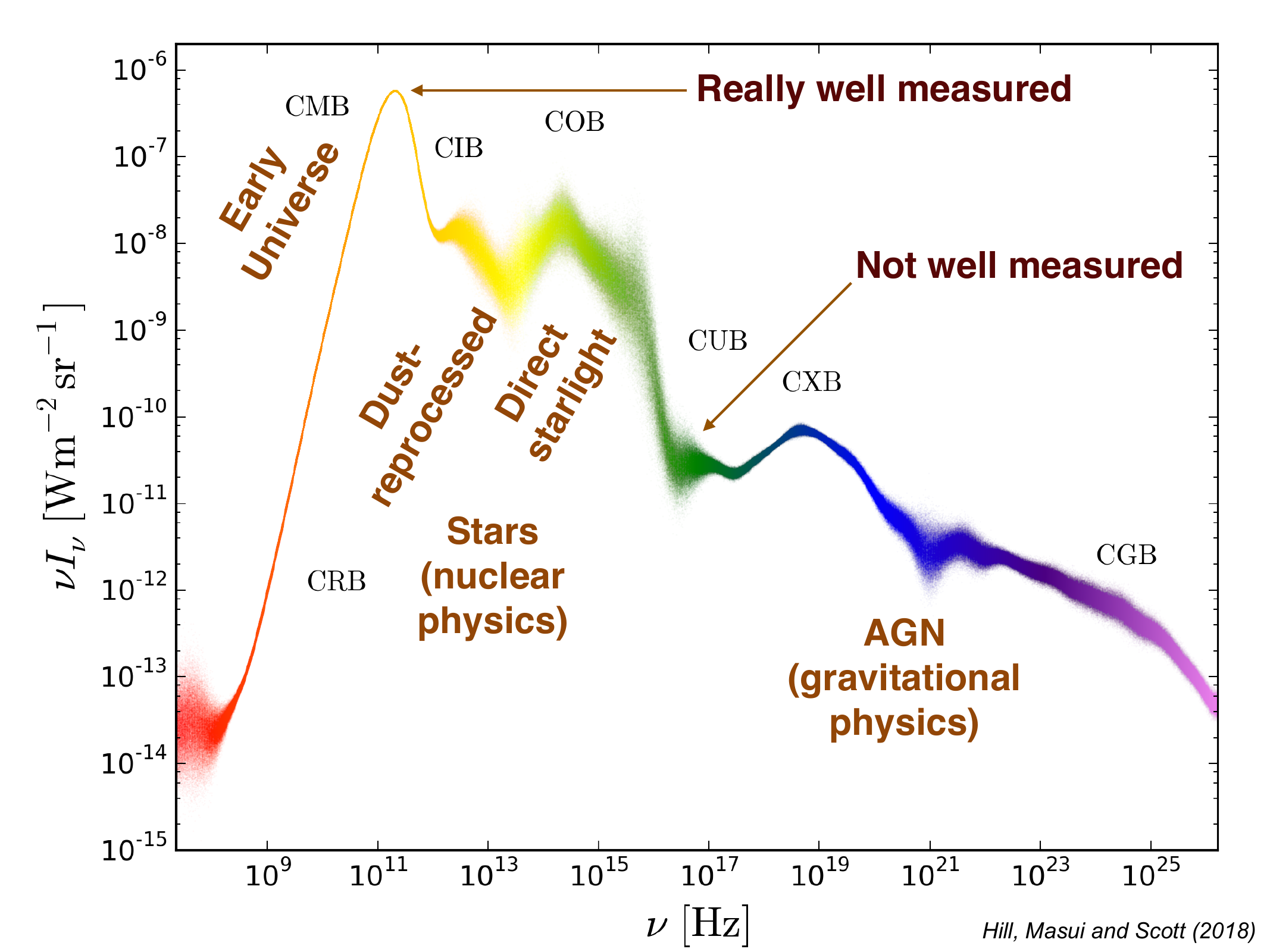}
\caption{Recent summary of constraints on the extragalactic photon background,
the ``spectrum of the Universe'',
with the width of the line indicating how uncertain the measurements are.
The vertical axis here is intensity times frequency, which gives the
contribution per logarithmic interval to the total intensity, and is also
proportional to the energy density.}
\label{fig:SoftheU}
\end{figure}

The origin of various parts of the photon spectrum
is labelled on a recent compilation of the ``Spectrum of the Universe''
by Hill, Masui \& Scott \cite{HillMS} in
Fig.~\ref{fig:SoftheU}.  This ``splatter'' diagram combines
data in each wavelength range, making a wide splash of colour where the
uncertainties are large and a thin line where the data are more precise
\cite{rainbow}.  A similar exercise can be performed for extragalactic neutrinos
-- except of course the data are mostly non-existent.  Still, we can estimate
what the cosmic neutrino spectrum would look like.  The best attempt to do this
is in a review by Vitagliano, Tamborra \& Raffelt \cite{Vitagliano}.  These
authors summarise backgrounds coming from sources on the Earth, the Sun, our
Galaxy and distant galaxies, in addition to the CNB.

%\vspace{0.5cm}

\section{\boldmath$\nu$ background spectrum}
\label{sec:nu_spectrum}
The CNB is similar to the CMB.  One important difference is that neutrinos
obey Fermi-Dirac instead of Bose-Einstein statistics.  That means that although
the spectrum is thermal, the distribution function in the spectrum has the form
$\{\exp(E/kT)+1\}^{-1}$ rather than $\{\exp(E/kT)-1\}^{-1}$.  Another
difference is that the neutrino temperature is $1.9454\,$K, rather than the
$2.7255\,$K of the CMB \cite{fixsen2009}.
This comes from the fact that as the Universe
expanded and cooled, the cosmic neutrinos decoupled {\it before\/} the last
particle annihilation, namely that of electrons and positrons.  As the
temperature dropped below the electron rest mass, the CMB was boosted by the
extra annihilation photons.  By comparing entropy densities before and after
this annihilation process, the ratio of neutrino to photon temperatures can be
shown to be $(4/11)^{1/3}$ \cite{picky}.

Another important difference between the CNB and CMB
is that because neutrinos have mass, they are
no longer completely relativistic.  The results of neutrino oscillation
experiments tell us that at
least two of the neutrino flavours will be non-relativistic by the present
time.  So how does this affect the spectrum?

For photons in an expanding universe, the temperature scales with redshift as
$T(z)=T_0(1+z)$ and the frequency as $\nu(z)=\nu_0(1+z)$,
so that the factor $\{\exp(h\nu/kT)-1\}^{-1}$ retains its shape at all
redshifts.  This is true whether you consider the frequency or the energy or
the momentum to be the quantity that redshifts (because for photons, all
these quantities are the same, up to physical constants).
However, for neutrinos, it turns out that it's really the
momentum that redshifts and so $\{\exp(pc/kT)+1\}^{-1}$ keeps its shape
\cite{picky2}.

\begin{figure}[htbp!]
\begin{center}
\includegraphics[width=0.8\textwidth]{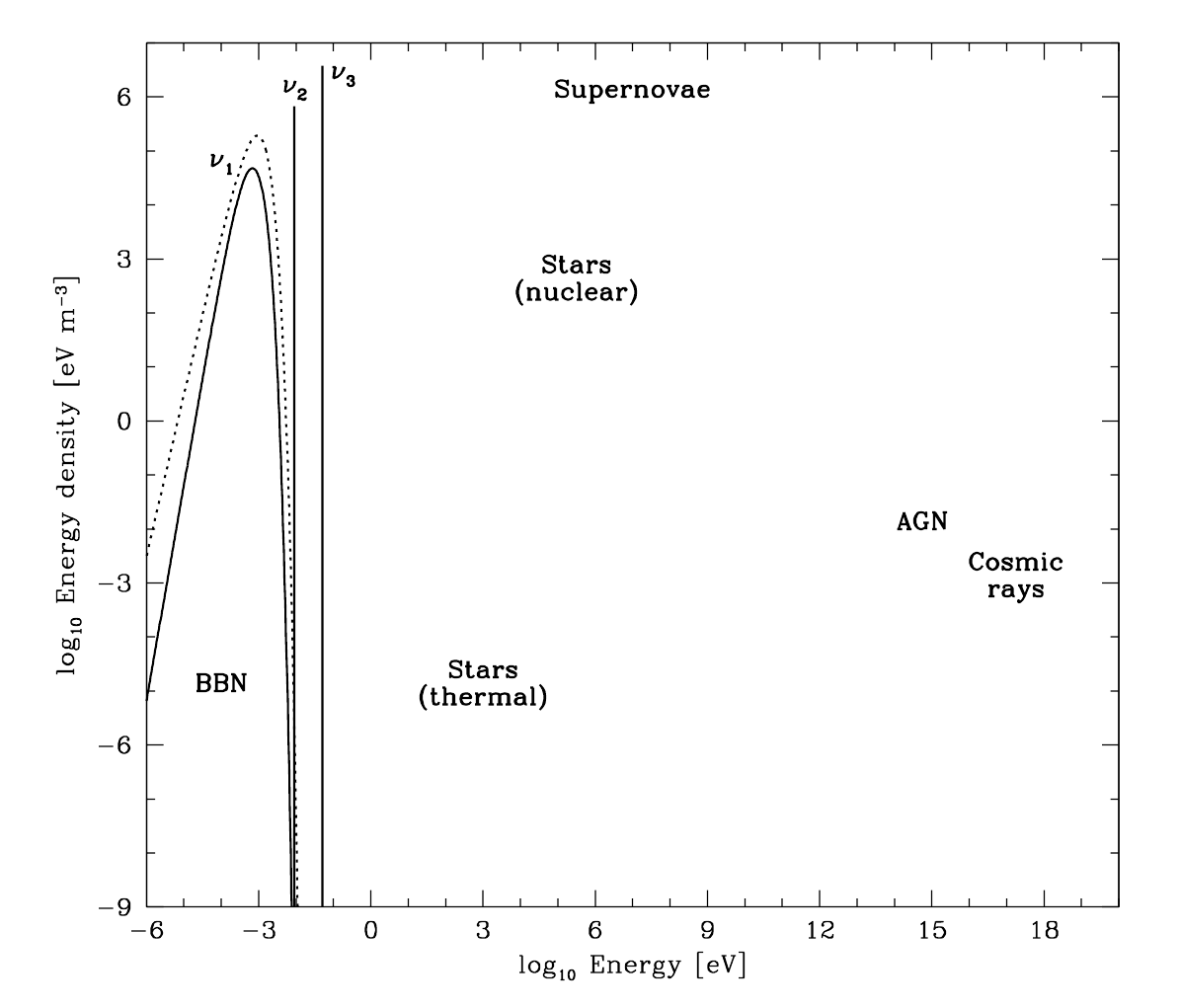}
\caption{Extragalactic neutrino backgrounds.
The dotted line is the CMB, represented as energy density per unit
energy, multiplied by energy.  In a similar way, the solid curves show the
CNB (for a minimal set of neutrino masses).
The other labels indicate the approximate locations in this plane of several
other contributions to the extragalactic neutrino background.  Apart from
perhaps at the highest energy end, this plot is essentially devoid of direct
data constraints.}
\label{fig:GUNS}
\end{center}
\end{figure}

What this implies is that if we plot the CNB as a function of momentum, then
all three neutrino flavours will have the same shape, {\it but\/} if we plot
as a function of energy, then things are more complicated.  Since
$E^2=p^2c^2+m^2c^4$, then for massive enough neutrinos, the energy will be
dominated by the neutrino's rest mass.  This is indicated in
Fig.~\ref{fig:GUNS} for the minimal model consistent with neutrino
oscillation results: $m_1\simeq0$; $m_2\simeq9\,$meV; and $m_3\simeq50\,$meV.
There are three separate CNB curves on the figure, with the two massive
neutrino states being essentially spikes, where the energies pile up at the
rest masses \cite{units}.

Figure~\ref{fig:GUNS} gives a broader representation of the neutrino
version of Fig.~\ref{fig:SoftheU} \cite{passthrough}.  It's plotted in units
of energy density per unit energy times energy (like for the photon
background, approximately showing contributions to energy density).  As well
as the CNB, there are other extragalactic neutrino sources.  There will be
neutrino spectral features coming from big-bang nucleosynthesis, in the same
way that the CMB spectrum will contain weak line features from atomic
recombination (e.g.\ Ref.~\cite{Wong2006}).  The Sun is
a locally dominant source of neutrinos, giving both thermal and nuclear
emission features -- stars in our Galaxy will contribute neutrinos at
essentially the same energies (roughly MeV), but with perhaps $10^8$ times
lower flux, while extragalactic stellar neutrinos will give a density another
couple of orders of magnitude weaker \cite{stars}.
These neutrinos will be an average over the spectra
from all types of star, with the widths of nuclear neutrino lines being
broadened by the redshift spread of distant galaxies.  The extragalactic
neutrinos will therefore give a spectrum like the Solar one, but about
$10^{10}$ times lower in amplitude, redshifted in energy by about a factor of
2 and smeared by the width of
$\Delta z\sim1$ for star-forming galaxies.  If we could detect them they
would give a direct measure of the history of star formation, with no
correction needed for interstellar extinction (like the case for photons).

There are also neutrino backgrounds expected from distant
supernovae (sometimes called the ``diffuse supernova background''),
from AGN (perhaps already partly being detected by IceCube)
and from cosmic rays (also called ``cosmogenic'' neutrinos).
The positions of each of these background sources are represented
approximately in Fig.~\ref{fig:GUNS} -- so this figure is really a cartoon
compared with the analogous photon-background plot.

%\vspace{0.5cm}

\section{\boldmath$\nu$ mixing}
\label{sec:mixing}
A fundamental difference between neutrinos and photons is that there are
three flavours of neutrino, but a photon is just a photon.  However, it's even
more complicated than this because the flavour states can mix into each
other \cite{timbits}.  Mass states aren't flavour states and the mixing matrix
is far from being diagonal.  In other words, it's hard to answer a question
like ``what's the mass of the electron neutrino?''

Neutrinos interact in flavour states, but propagate in mass states and if CNB
detectors existed, they would presumably be sensitive to particular flavour
states.  That means
that the neutrinos in the CNB decouple in the early Universe as $\nu_e$,
$\nu_\mu$ and $\nu_\tau$, but propagate to us as $\nu_1$, $\nu_2$ and $\nu_3$.
If we could easily
detect them, then we'd have to consider whether we were detecting, say,
just $\nu_e$s or all flavours, and we'd have to take the mixing into account.
For example, with what we know about the mixing matrix, an experiment that
detects electron neutrinos from the CNB will be detecting approximately
68\,\% of the smallest mass state ($m_1\simeq0$), 30\,\% of the next mass state
($m_2\simeq9\,$eV in the normal hierarchy) and only 2\,\% of the most massive
state ($m_3\simeq50\,$meV).

\section{\boldmath$\nu$ speeds}
\label{sec:speeds}
The CNB neutrinos were in equilibrium with other particles until the
weak interactions froze out at $t\sim1$\,s in the history of the Universe.
At those early times the neutrinos had energies that were much higher than
their rest masses and hence they were ultra-relativistic.  Their momenta
redshifted according to $p\propto(1+z)$, which means that the Lorentz factor
$\gamma\propto(1+z)$ until $\gamma\simeq1$ (so $v\simeq c$), after which
$v\propto(1+z)$.  The transition happened when momentum was approximately
rest mass (times $c$).

In the simplest scenario, the two heavier mass-state neutrinos are now
non-relativistic, while the lightest mass state is still relativistic.
The average speed of cosmic neutrinos, after they become
non-relativistic is
\begin{equation}
\left\langle v_\nu\right\rangle
 \simeq150\,\left({{\rm eV}\over m_\nu}\right)\,{\rm km}\,{\rm s}^{-1}
\end{equation}
(e.g.\ Ref.~\cite{Abazajian2016}).
This means that for a 50-meV mass state (probably the heaviest in the normal
hierarchy), the transition to the non-relativistic regime happened at
$z\simeq100$ and the average speed is now about
$3000\,{\rm km}\,{\rm s}^{-1}$, i.e.\ only 1\,\% of the speed of light.  If the
second mass state is around 9\,meV, then those neutrinos are travelling about
6 times faster than the heaviest neutrinos,
while the lowest mass state is still relativistic today.
If the hierarchy is inverted, then slower speeds could prevail.

\section{\boldmath$\nu$ last-scattering surface(s)}
\label{sec:LSS}
Cosmic neutrinos decoupled in the early Universe when the rate of
weak interactions, like $p+e^{-} \leftrightarrow \nu_e+n$,
became too slow compared with the expansion rate.  This occurred when
average particle energies were around 1\,MeV, at $t\sim1$\,s and redshift
$z\sim10^{10}$.
We can thus think of a ``neutrino last-scattering surface'' (LSS) in analogy
with the CMB LSS, which happens at $t\simeq380{,}000$, $z\simeq1100$.

However, the last section described how
some neutrinos have been non-relativistic for
quite some time.  As first pointed out by Bisnovatyi-Kogan \& Seidov
in 1983 \cite{Bisnovatyi-Kogan} (and described more explicitly by Dodelson \&
Vesterinen \cite{Dodelson2009}), the finite speed of neutrinos means that it
is possible for the {\it distance\/} to the neutrino LSS to be significantly
{\it smaller\/} than the distance of the photon LSS, even though the
neutrinos decoupled from the rest of the matter much earlier than the photons
did \cite{neutrino_cone}.  Neutrinos aren't on the light cone, but
instead they live on the
neutrino cone (which makes an angle $<45^\circ$ with the time axis).
As shown by Dodelson \& Vesterinen, for neutrino masses greater than about
0.1\,meV, the CNB LSS will be closer than the CMB LSS.

The situation is made more interesting because of the multiple flavours of
neutrino and the mixing of those flavours.  There will actually be three
distinct LSSs, one for each mass state, since those have different average
speeds.  In
more detail, the distance will be different for every neutrino momentum in the
Fermi-Dirac distribution, and hence each LSS will be quite wide in distance.
With a detector, say of $\nu_e$s, a fraction will come from each of
the broad LSSs corresponding to the three mass states \cite{sadly}.

The fact that neutrinos live off the light cone gives the potential prospect
of seeing the same cosmological structure at two different epochs, one using
neutrinos (at an earlier time) and the other using photons (at a later time)
\cite{multi}.
We could imagine some overdense region where varying gravitational
potential (``ISW'') or lensing effects might be detectable in both photons and
neutrinos.  Then one could directly measure the growth of the overdensity
between the two epochs, yielding a new way to constrain the behaviour of the
fluids affecting the growth, e.g.\ the equation of state of the dark energy
\cite{growth}.

\section{\boldmath$\nu$ anisotropies}
\label{sec:anisotropies}
The CMB contains anisotropies arising from density perturbations in the early
Universe \cite{ScottSmoot},
and this will also be true for the CNB.  The first publication of a
prediction of the CNB anisotropy power spectrum was in a 1995 paper by
Hu, Scott, Sugiyama \& White \cite{HSSW}.  The relevant plot \cite{HSSWnote}
showed that
there would be an enhancement in the $C_\ell$s at about the CMB LSS horizon
size because of the ``early ISW'' effect, raising the power spectrum to a
plateau that would continue out to the scale of the CNB LSS, with damping at
multipoles $\ell\sim10^8$ (i.e.\ milli-arcsecond angular scales).

\begin{figure}[htbp!]
\begin{center}
\includegraphics[width=\textwidth]{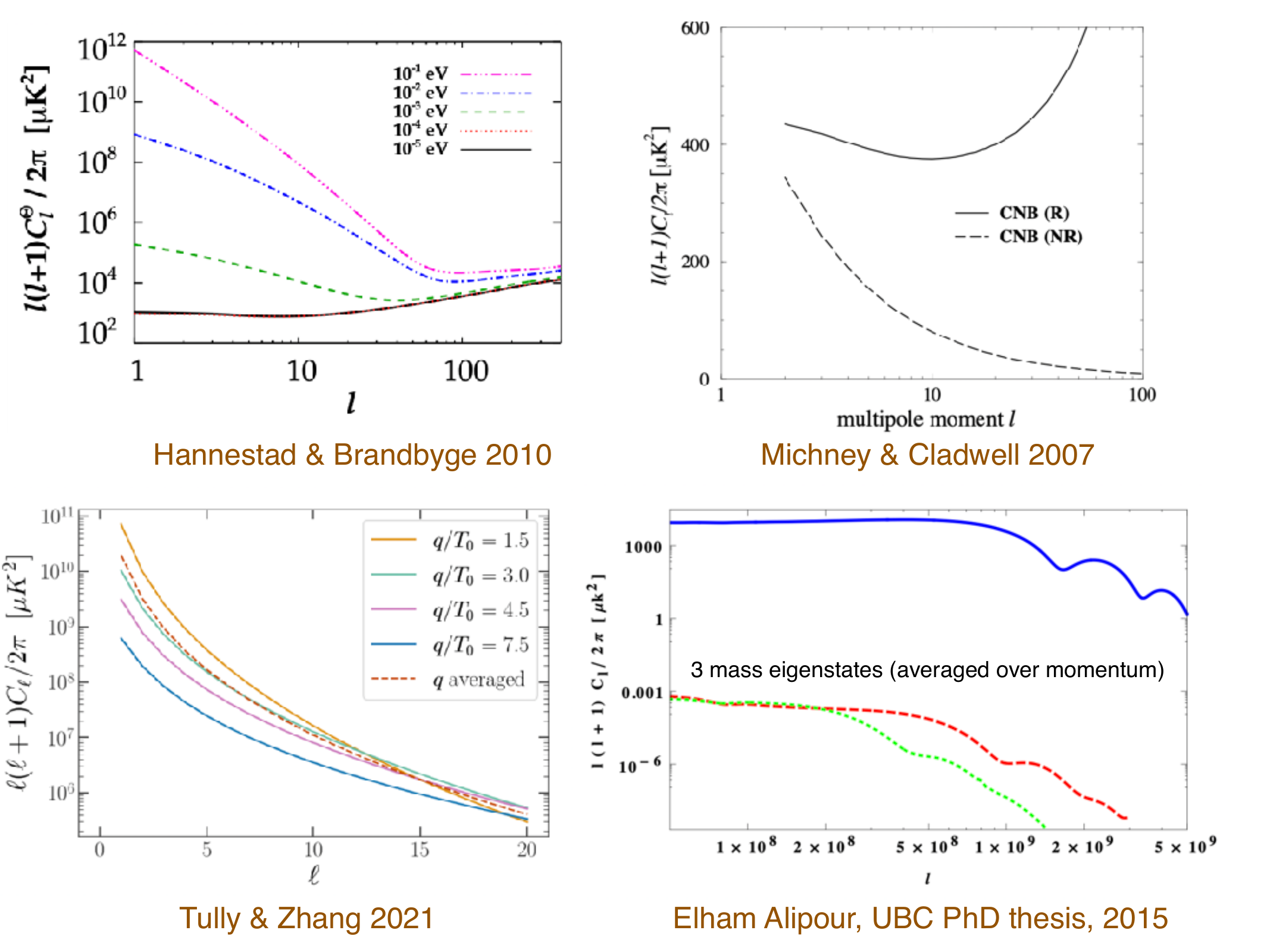}
\vspace{-0.5cm}
\caption{Some published calculations on CNB anisotropy
(actually almost all of them).}
\label{fig:anisotropies}
\end{center}
\end{figure}

This initial calculation ignored the ``late ISW'' effect that gives an
extra contribution at low multipoles (as it does for the CMB), because of
gravitational potentials changing fairly recently in the history of the
Universe.  Examples of more accurate calculations
from some more recent theoretical studies are shown in
Fig.~\ref{fig:anisotropies}.  Mostly these have focussed on the low-$\ell$
regime \cite{Michney2007,Hannestad2010,Tully2021}.  At higher
multipoles, the most detailed analysis is in the 2015 Ph.D.\ thesis of
Elham Alipour, supervised by Kris Sigurdson at UBC \cite{Elham}.  The
calculations are complicated to get right, and also computationally
challenging.  That's because each momentum state in the neutrino distribution
function has to be evolved separately -- different momenta travel at different
speeds after all!  This means that when running a Boltzmann solver, in addition
to following sets of $k$ (Fourier-space) modes and spherical harmonic (angular)
modes, there also has to be a loop sampling over the momentum space (which
is of course unnecessary for photons, all of which propagate at the speed of
light).

%\vspace{0.5cm}

If we could figure out a way to measure the CNB anisotropies, then they would
in principle yield astonishingly precise cosmological constraints.
The information content
of a 2-d Gaussian field, like the CMB sky, is just a matter of mode counting,
with the number of modes out to a maximum multipole $\ell_{\rm max}$
scaling as $\ell_{\rm max}^2$ \cite{Scott2016}.  The number of available modes
in the CNB is thus a factor of around $10^{10}$ larger than for the CMB.  The
CNB power spectrum is fairly flat out to high $\ell$, lacking the deep
oscillatory structure in the CMB power spectrum, so the CNB anisotropies
would not necessarily improve all cosmological parameter constraints by such a
large amount, but nevertheless, the resulting uncertainty on some parameters
(e.g.\ the scalar spectral index, $n_{\rm s}$) would be extraordinary.
And, as shown by Alipour,
oscillatory structure in the damping tail (at $\ell\sim10^9$) would
likely yield further parameter constraints \cite{far-fetched}.

\section{\boldmath$\nu$ inhomogeneities}
\label{sec:inhomog}
Treating CNB anisotropies in analogy with CMB anisotropies isn't quite the
right thing to do though.  That would be fine if cosmic neutrinos were
still completely relativistic, but they're not.
At the other (non-relativistic) extreme, we have
CDM particles, which are ``cold'' today, meaning that their
typical velocities are negligible for the purposes of structure formation.
The CDM sky has no anisotropies that are remnants of the epoch when those
particles decoupled in the early Universe, since the trajectories have been
dramatically altered by interaction with gravitational potentials.  Instead of
primordial anisotropies on the sky, CDM particles exhibit spatial
clustering, i.e.\ inhomogeneities, with enhancements in potentials and
streams from halo interactions as structures merge together.

There have been quite a few studies of the clustering of cosmic neutrinos,
either through analytical calculations or
treating them as particles in simulations,
\cite{Singh2003,Ringwald2004,Paco2013,Mertsch2020,Yoshikawa2020,Elbers2023,Zimmer2023}.
These analyses look at how neutrinos might be enhanced in the halos
of galaxies, or perhaps in the Solar neighbourhood of our own Galaxy.
The consensus view is that effects of neutrino enhancement in galaxy
halos are relatively weak, at perhaps the few percent level.

Cosmic neutrinos lie somewhere (in their behaviour)
between photons and CDM particles.  So, to
fully treat their fluctuations we need to consider both the primordial
anisotropies and also the changes caused by gravitational interactions while
they (or at least some of them) were non-relativistic.

These effects will include gravitational lensing.  The usual calculation of
the gravitational bending angle gives
\begin{equation}
\alpha = {4GM\over c^2 b},
\end{equation}
for a photon passing mass $M$ at impact parameter $b$.  Since this involves
the square of the speed in the denominator, then for
neutrinos moving at $v\ll c$, the deflections can be {\it much\/} larger than
for photons \cite{Patla2014}.
It was pointed out earlier that the highest mass neutrinos
have average speeds today that are about 100 times slower than $c$.  Hence a
deflection through a galaxy cluster that would bend photons by arcminutes could
completely change the trajectory of a slow neutrino.  Indeed, there will be
situations where the path is bent so much that the neutrino is captured by the
mass, which is when ``lensing'' turns into ``clustering''.

Some strong lensing effects for CNB neutrinos were investigated by Yao-Yu Lin
\& Holder \cite{LinHolder2020}, showing for example that the chromatic nature
of neutrino lensing means that it would be possible to probe the whole causal
volume.  These studies are separate from the related topic of
investigating {\it photon\/} lensing in halos affected by neutrinos
(e.g.\ Refs.~\cite{Paco2011,Hotinli2023}).  A full analysis of all anisotropy,
clustering and lensing effects in the CNB has yet to be done.

\section{\boldmath$\nu$ dipoles}
\label{sec:dipoles}
The largest scale anisotropy in the CNB will be the dipole.  Just as is the
case for the CMB, the dipole is expected to be dominated by our motion
relative to the universal ``rest frame'' (e.g.\ Ref.~\cite{Sullivan2022}).
The dipole can also been measured
in the anisotropy of the distribution of sources on the sky, such as galaxies
or AGN.  If these sources are close enough then there will also be a dipole
component caused by source clustering.  Inhomogeneities in the cosmic
neutrinos could therefore affect the size and direction of the dipole.
For the CNB, the dipole could (at
least in principle) be measured for different flavours, coming from different
combinations of mass states, and this could be done for different neutrino
momenta, probing different volumes.

There have been papers suggesting related tests.  For example,
there is a possibility of using the range of low-order neutrino
anisotropies as a function of momentum to
perform a test of the Copernican principle \cite{Jia2008}.  And the annual
modulation of the CNB could also potentially be used to extract
the cosmic neutrino signal \cite{Safdi2014,Huang2016,Akhmedov2019} -- this is
essentially using the time-variation of the dipole a bit like the ``orbital
dipole'' is used to calibrate CMB measurements.

\section{\boldmath$\nu$ indirect detection}
\label{sec:constraints}
If the CNB is so hard to detect, why are we so confident that it exists?  The
most important reason is that there's strong (if indirect) evidence of the
CNB through the precision measurements of CMB anisotropies.  The neutrinos
contribute at early times to the relativistic background (or ``radiation''),
affecting the expansion rate to such an extent that the CMB data cannot be
fit without including the neutrinos (which give 68\,\% extra energy density
compared to the
photons).  Since matter-radiation equality happens not much earlier
than the cosmic recombination epoch, then radiation is still important at the
recombination time and in fact $\rho_\nu\sim0.1\rho_{\rm tot}$ at the CMB
LSS, making it clear why the CNB is far from negligible.

The strength of
this effect can be assessed through the effective number of light degrees
of freedom, $N_{\rm eff}$.  For three neutrino flavours we would expect
$N_{\rm eff}=3$, except that detailed calculations, including non-instantaneous
decoupling, oscillation effects, finite-temperature effects, etc.
\cite{Bennett2021}, give $N_{\rm eff}=3.044$ \cite{coincidence2}
as the currently best accepted value
(with theoretical uncertainties being in the last digit).

The Planck data, including CMB lensing, combined with baryon acoustic
oscillation (BAO) measurements from galaxy surveys
(see Ref.~\cite{planck2016-l06} for details and references) gives
\begin{equation}
N_{\rm eff}=2.99\pm0.17.
\end{equation}
This effectively corresponds to a $17\,\sigma$ detection of the CNB.

In addition, the neutrinos should have two parameters that describe the
behaviour of their perturbations, $c_{\rm eff}$, the sound speed, and
$c_{\rm vis}$, which parameterises the anisotropic stress.  Both are expected
to have the same value for non-interacting massless neutrinos, namely
$c_{\rm eff}^2=c_{\rm vis}^2=c^2/3$.  Treating these as additional free
parameters and fitting to the Planck data yields values of
$0.324\pm0.006$ and $0.327\pm0.037$, both consistent with expectations, and
an anisotropic stress of zero is rejected at the $9\,\sigma$ level
\cite{planck2014-a15}.  A related effect is a phase shift in the
CMB power spectra, caused by the anisotropic stress,
which has also be measured \cite{Follin2015}.

To summarise these results: the CNB {\it has\/} been detected, indirectly,
through its effect on CMB anisotropies.  We know that before the recombination
epoch there were three species of relativistic particles, which had the same
properties as expected for neutrinos.  Because of the strength of this
evidence, the standard $\Lambda$CDM model includes the CNB, with fixed
properties.

The standard cosmological model (SMC \cite{SMC2006,skeptics})
is $\Lambda$CDM, with a
parameter space spanned by six parameters, e.g.\ $\Omega_{\rm b}h^2$,
$\Omega_{\rm c}h^2$, $\theta_\ast$, $A_{\rm s}$, $n_{\rm s}$ and $\tau$.
The temperature of the CMB would be a 7th parameter, but this is so well
determined that it is usually considered to be fixed, rather than being
allowed to float when fitting power spectra (but see Ref.~\cite{Wen2021}).
The SMC isn't like the standard model of particle
physics, which has a clear mathematical basis and hence a fixed amount of
freedom.  Most cosmologists are expecting that eventually we'll need more
parameters to describe the SMC, hopefully with some genuine surprises.
The next parameter that we think we're about to require is related to
neutrino mass.

\section{\boldmath$\nu$ mass}
\label{sec:mass}
The sum of the neutrino masses is usually written $\sum m_\nu$.  But an
expression is an awkward notation to use for a parameter,
and it would be better called something such as ${\cal M}_\nu$.  
The Planck Collaboration used a basic cosmological model that includes the CNB
with a single mass eigenstate having $m_\nu=0.06\,$eV.  But it would make no
difference to the fit if the mass was split
between two or three neutrinos -- the sensitivity
is essentially to ${\cal M}_\nu$ alone, and not how the mass is distributed
among types of neutrino.

The tightest constraints from the CMB alone come when including CMB lensing,
giving numbers like ${\cal M}_\nu<0.25\,$eV at 95\,\% confidence from the
Planck 2018 data set \cite{planck2016-l06}.  Combining with BAO data yields
even stronger constraints.  Currently the tightest limit comes from using the
newest analysis of the ``Public Release 4'' Planck data \cite{Tristram2023} in
combination with BAO results, yielding
\begin{equation}
{\cal M}_\nu<0.11\,{\rm eV}\ (95\,\%)\ \cite{2dplane}.
\end{equation}
This is coming very close to ruling out the inverted mass hierarchy, which
requires ${\cal M}_\nu\simgt0.1\,$eV (e.g.\ Ref.~\cite{Jimenez2022}).

Right now the evidence for the existence of the CNB does not require the
neutrinos to have any mass.  We need to include the CNB to fit the Planck
satellite data, but a fully relativistic (i.e.\ massless)
CNB is sufficient for now.  However,
that situation is expected to change soon, as cosmological data become
increasingly precise.

There are three separate effects of neutrino mass on the CMB, all of which
contribute to current constraints on ${\cal M}_\nu$:
\begin{enumerate}
\item changing the distance to the CMB LSS;
\item smoothing of power spectra at small scales;
\item changing the shape of the lensing power spectrum.
\end{enumerate}
This last item is particularly important, especially as ground-based CMB
experiments push to higher sensitivity and angular resolution, and is
expected to be the cleanest way of detecting neutrino mass.  Additionally,
massive neutrinos change the shape and growth history of the matter power
spectrum (relative to the CMB amplitude), although the effect occurs on
small scales, where the power spectrum becomes nonlinear.

Tight constraints are expected to come from combining the CMB with surveys
that measure the expansion rate at low redshifts, e.g.\ from the BAO scale.
The Euclid satellite is expected to achieve a $1\,\sigma$ uncertainty on
${\cal M}_\nu$ of around 30\,meV \cite{Euclid}.  The Dark Energy Spectroscopic
Survey Instrument  \cite{DESI} and Rubin/LSST are similar \cite{LSST}.
Future CMB experiments such as
CMB-S4 will also provide important constraints through the effect on CMB
lensing; better determination of
the reionisation optical depth, such as provided by the LiteBIRD 
satellite \cite{LB}, will help break parameter degeneracies to
enable improved ${\cal M}_\nu$ determination.  Overall there is an expectation
that an uncertainty of around 20\,meV is possible
(e.g.\ \cite{Dvorkin,PanKnox,Mishra-Sharma}), perhaps even as small as
10\,meV if all constraints are combined.

\section{\boldmath$\nu$ exotica}
\label{sec:exotic}
Theorists are creative, and hence there are many ideas for extending the
physics of neutrinos to make things more complex.
One commonly discussed addition to the neutrino family is to consider
an extra type that has essentially no interactions, and hence is
called a ``sterile'' neutrino \cite{Abazajian2017}.
This could have a different distribution function than the
normal CNB types and hence contribute as a fraction (rather than an integer)
to the value of $N_{\rm eff}$.  An extra (fourth) kind of neutrino would
involve additional mixing matrix parameters, such as $\sin^22\theta_{14}$,
and squared mass-splittings, such as $\Delta m_{41}^2$.
The two-dimensional space of $(\sin^22\theta_{14},\Delta m_{41}^2)$ can be
related to the cosmological parameters $({\cal M}_\nu,N_{\rm eff})$.  However,
one needs to be careful with volume effects, sampling and priors, in order to
properly transform between the particle-physics and cosmological parameter
spaces \cite{Bridle2017,Knee2019}.

There are plenty of other ideas that could change the physics of the CNB.
Perhaps the least speculative is the possibility that there might be a
non-zero chemical potential in the neutrinos, or other changes to the
Fermi-Dirac spectrum.  They could exhibit decays, annihilations,
self-interactions, other non-standard interactions, Lorentz violation or
additional kinds of oscillation.  There could also be new sources of low-energy
neutrinos that complicate the interpretation of the CNB.  Much more could
be said about these and other possibilities, but we shall stop here.

\section{\boldmath$\nu$ conclusions}
\label{sec:conclusions}
The CNB undoubtedly exists, with clear evidence of its early-time effects on
the CMB, when still part of the relativistic background.  The inferred count
of relativistic species is consistent with the number three \cite{three},
and the behaviour of the fluid perturbations is also in line with what we
expect for the CNB.

Current cosmological data place a limit on the sum of the neutrino masses,
${\cal M}_\nu$, which is only a factor of 2 or so from the minimum expectation
from neutrino oscillation experimental results.
The inverted hierarchy is already close to being ruled out
(or confirmed for that matter) by the data.
Cosmologists are now excited about the possibility of actually constraining
${\cal M}_\nu$ to be non-zero through the effects of
neutrino mass on the power spectra of CMB anisotropies (particularly lensing)
and galaxy clustering.
Combining all the measurements expected in about the next decade should enable
a measurement of ${\cal M}_\nu$ at around the $5\,\sigma$ level, even for
the minimal mass limit in the normal hierarchy.

\begin{figure}[htbp!]
\begin{center}
\includegraphics[width=0.8\textwidth]{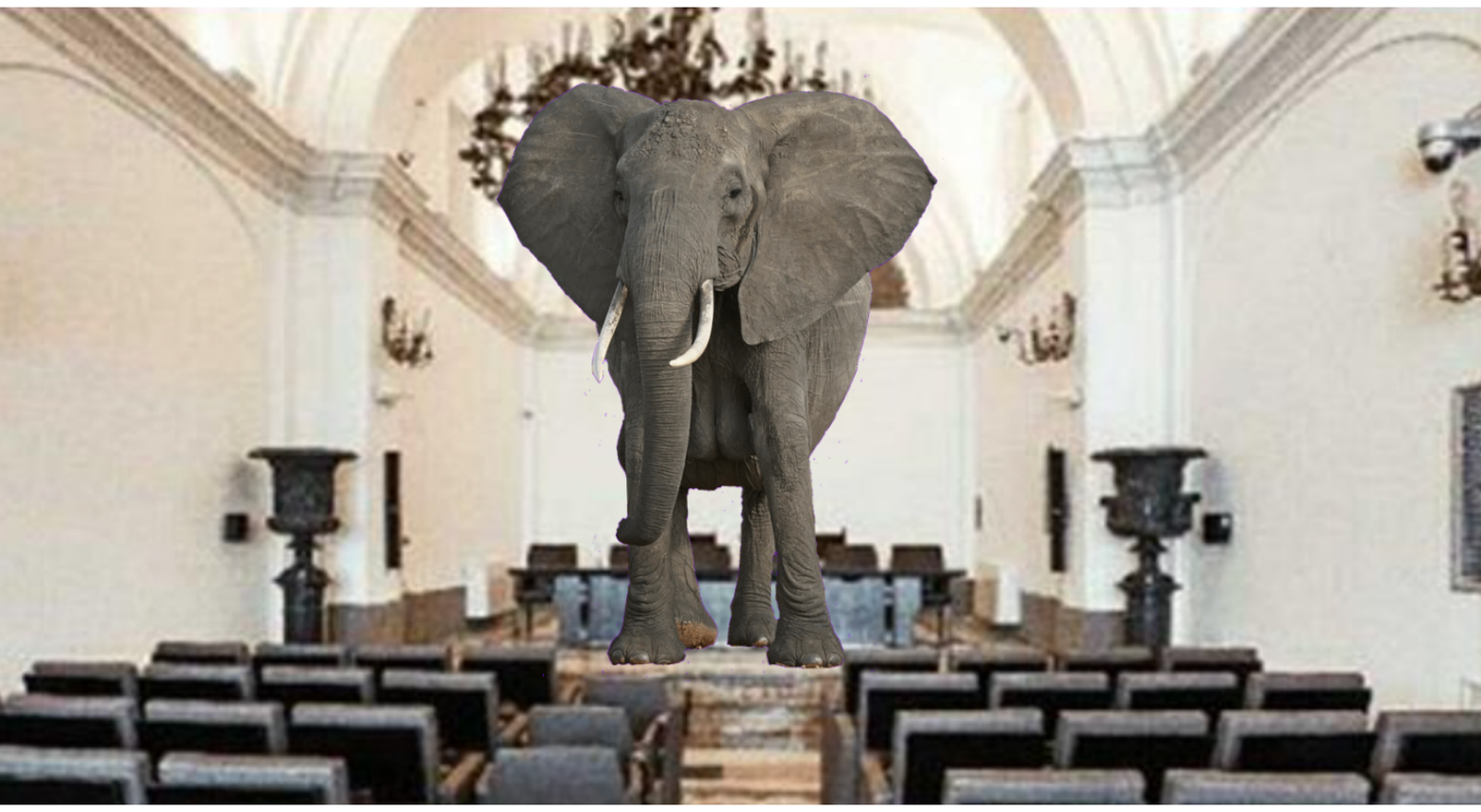}
\vspace{-0.5cm}
\caption{The elephant in the room.  In this case, the Fermi Room, where the
lectures of this summer school were given.}
\label{fig:FermiElephant}
\end{center}
\end{figure}

However, there's an elephant in the room (Fig.~\ref{fig:FermiElephant}).
There are many particle experimentalists working hard to try
to detect the neutrino mass directly and the question is -- will they take
seriously a cosmological detection of ${\cal M}_\nu$?
It seems that the chasm between the two sub-fields has been closing,
and nowadays both particle and astro researchers tend to
appreciate the complementarity of the approaches.  The cosmological
measurement of neutrino mass might be indirect and ``model-dependent'', but the
model is understood to be really a very simple one that already fits the data
astonishingly well.  There are multiple kinds of cosmological observable that
can be affected by massive neutrinos, and if two or more start to point towards
some value of the total neutrino mass, then that will have to be taken very
seriously.  Of course it will still be useful to make more direct
measurements of neutrino mass, as well as to determine what each of the
individual masses are.  And on the flip side, although future
particle-physics experiments might be able to reach down to the neutrino
mass scale, they are very very far from being
able to detect the neutrinos in the CNB itself.

Perhaps through the pursuit of increasingly sensitive ways of probing
low-energy neutrinos, we might eventually reach a situation where the CNB can
be observed directly, enabling us to probe cosmological perturbations at the
neutrino decoupling epoch.  It's not going to happen soon, but we can dream!

%\vspace{0.5cm}

\acknowledgments
I would like to thank the organisers for an entertaining school in Varenna,
and the students and other participants for interesting discussions.

\bibliographystyle{varenna}
\bibliography{varenna_CNB}

\begin{thebibliography}{10}
\bibitem{summary}
If you just want a summary, you don't have to read any further.
\bibitem{simgt}
For example in the first list above I'm using the ``$\simgt$'' symbol in the
  physics way, meaning ``bigger than but about the same as'', and not the
  astronomy way, ``bigger than something that's only known approximately''.
\bibitem{minimise}
Depending on many factors, so this number is quite approximate. You can
  minimise the number by always trying to orient your body along the line of
  sight to the Sun, keeping your cross-section as small as possible.
\bibitem{undetectable}
There have been several proposals for ways to try to detect the CNB directly.
  However, even the most optimistic proponent of any of these ideas would agree
  that it will be challenging, and a positive result is not likely in the near
  future.
\bibitem{coincidence1}
Unless you can think of a physics reason for this equivalence.
\bibitem{CDM}
When faced with colleagues who seem to have difficulty believing in the
  existence of CDM, i.e.\ an almost non-interacting but ubiquitous particle
  whose existence is supported by a mountain of empirical data, I often wonder
  if they believe in neutrinos either!
\bibitem{SM}
In fact Ted Ressell kindly passed along his supermongo scripts (for those who
  remember what that means!).
\bibitem{rainbow}
A joke popularised by supernova-expert Bob Kirshner is that the definition of a
  spectroscopist is someone who takes a rainbow and makes it into a graph; here
  we have done the opposite, taking a graph and turning it into a rainbow.
\bibitem{picky}
It may be worth mentioning here that there are roughly 1\,\%-level effects
  coming from the fact that there are three flavours of neutrinos that don't
  decouple instantaneously, or at quite the same time. This means that the
  third digit in the constants to use for the neutrino temperature, number
  density, energy density, etc., have to be considered carefully.
\bibitem{picky2}
To be more picky, the neutrino spectrum isn't exactly thermal, because of the
  fact that neutrinos with higher momentum receive more energy during pair
  annihilation. The spectral distortions in the CNB (like for the CMB) would
  provide additional constraints on early Universe physics.
\bibitem{units}
I often tell students that one of the hardest things to get right in a
  calculation is the units, and this is a good example! It would be simplest to
  plot the CNB in momentum units, where the three spectra are the same, but
  that doesn't give a figure that's very analogous with the photon spectrum.
  One option would be to stick with differential units, but calculate the
  energy (or momentum or just number) flux per unit energy or momentum. In an
  effort to plot something like energy density, we can calculate the energy
  flux per unit energy and multiply by one more power of energy, which is like
  performing an integral for a broad spectral feature -- but this fails to
  represent the correct situation for any line-like features (like the massive
  neutrino parts of the CNB). Hence the figure remains only a kind of cartoon.
\bibitem{passthrough}
Although, unlike the situation for photons, since neutrinos pass through
  essentially everything, it's always going to be hard to discriminate nearby
  (terrestrial, Solar system or Galactic) neutrinos from distant ones!
\bibitem{stars}
This estimate comes from assuming that stellar neutrinos are approximately
  proportional to optical luminosity and scaling by the ratio of the amount of
  light coming to us from the Sun relative to the cosmic optical background
  (COB).
\bibitem{neutrino_cone}
This is quite a surprising result for traditional cosmologists to wrap their
  heads round, since (almost) everything in cosmology is observed using
  photons, and photons always live on the light cone -- so you are always
  observing things the light travel time ago. But not so for non-relativistic
  neutrinos.
\bibitem{sadly}
Sadly, a detector of $\nu_e$s (surely the easiest to imagine buying at the CNB
  detector store) will only pick up a small fraction of the high mass (and
  hence slowest speed) neutrinos.
\bibitem{multi}
This would be ``single tracer'', but genuinely ``multi-messanger'' cosmology.
\bibitem{growth}
Don't expect this sort of measurement to be done soon though. Perhaps we'll
  have to wait so long that we'd already be able to see growth in real time!
\bibitem{HSSWnote}
The second author claims that the others didn't want to include the relevant
  figure, arguing that no one would care about measuring $10^{-5}$ fluctuations
  in an undetectable background; but eventually the other authors were worn
  down!
\bibitem{far-fetched}
Just to be clear, we're talking about variations of roughly 1 part in $10^5$,
  at angular scales below 1\,milli-arcsec, in an essentially undetectable
  background! Calculation of the effects of polarisation and gravitational
  waves are left as an exercise for the student.
\bibitem{coincidence2}
In the list of six best-fit cosmological parameters from Planck satellite data,
  the value of the scalar power-spectrum amplitude is $\ln(10^{10}A_{\rm
  s})=3.044$, which is another happy coincidence.
\end{thebibliography}

\end{document}